# The electronic activity of boron and phosphorus impurities in a-Si and a-Si:H

Bin Cai and D. A. Drabold
Department of Physics and Astronomy, Ohio University
Athens, OH 45701, U.S.A.

## ABSTRACT

In amorphous materials, acceptor and donor impurities rarely dope the system (shift the Fermi level). We find out *why* in a-Si:H. We report simulations on B and P doping of a-Si:H and a-Si. We analyze the Electronic Density of States (EDOS) with concentrations ranging from 1.6% to 12.5% of B or P in a-Si. The results indicate that tetrahedral B and P are effective doping configurations in a-Si, but high impurity concentrations introduce defect states. Clustered B or P also introduced mid-gap states. For a-Si:H, we report that both B(3,1) and P(3,1) (B or P atom bonded with three Si atoms and one H atom) are effective doping configurations. We investigate H passivation in both cases. There exists a "hydrogen poison range" for which H can modify the dopant configuration and suppress doping. For B doping, nearby H prefers to stay at the bond-center of Si-Si, leaves B four-fold and neutralizes the doping configuration; for P doping, nearby H spoils the doping by making tetrahedral P three-fold.

## INTRODUCTION

By introducing B or P, a-Si:H may be doped either n-type or p-type [1], a point of profound technological importance. In c-Si, B and P doping has been extensively studied. Because of translational invariance, impurities are compelled to have the same local tetrahedral environment as Si. According to the *8-N* rule, B atoms create a hole when they have $T_d$ symmetry, and P similarly donates an electron. The doping efficiency is almost 100%. In c-Si:H, H atoms passivate doping by relaxing the strain, and rendering B or P doping-inactive by enabling the impurities to become three-fold [1].

However, in contrast to c-Si, there is no periodic lattice in a-Si. The absence of a unique atomic environment leads to site-dependent doping as seen in studies with low concentration Boron[2]. However, theoretical studies on P doping and high concentration of B are still required. In a-Si:H, NMR[3] shows that, for B doping, 40% of the B has a nearby H at 1.6 Å; for P doping, 50% of P has a H at 2.6 Å. The doping efficiency is very low, and Boyce and Ready have conjectured that the sluggish doping may be due to H passivation[4]. But the atomistic mechanism of H passivation in doped a-Si:H is unclear.

In this paper, we report molecular dynamic simulations on B and P doped a-Si and a-Si:H, focusing on the electronic structure. For a-Si, we report the electronic density of states (EDOS) for various impurity concentrations. We attempt to find the effective doping and non-doping configurations. For a-Si:H, we mainly focus on describing the H passivation mechanisms with low concentration of impurities (<2.0%). By manually placing H in the models, we are able to investigate energetically preferred positions and explore consequent electronic structure.

## THEORY

All calculations were performed using the plane wave LDA code VASP with ultrasoft pseudopotentials and the local density approximation [5]. A previously generated defect-free 64-atom a-Si model was used as the initial configuration. B or P atoms were introduced into the network by substituting for Si atoms. Conjugate gradient relaxations were performed at constant volume. We obtained relaxed 1.6%, 3.1%, 7.8% and 12.5% B or P doped a-Si models. To study H passivation in B- or P-doped a-Si:H, we introduced H atoms at particular sites of the 1.6% B- or P-doped a-Si models with various distances from impurities.

## DISCUSSION

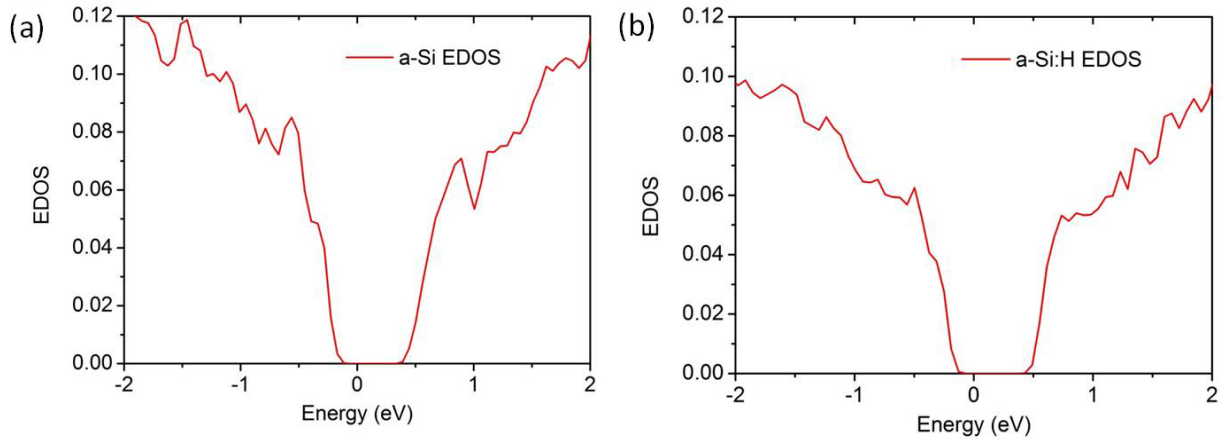

**Fig. 1** Electronic density of states (EDOS) (a) a-Si. (b) a-Si:H. The Fermi level is at 0eV.

We first present the electronic density of states of a 64-atom a-Si model and a 70-atom a-Si:H model with 8.5% H in Fig. 1. Both a-Si and a-Si:H models exhibit gaps unsullied by defect states. In the following, we first discuss B- and P-doped a-Si with impurity concentrations from 1.6% to 12.5%. Then, we investigate the mechanism of H passivation in both systems.

### Boron doped a-Si

We plot the EDOS of seven B-doping models [configuration (1)-(7)] in Fig. 2. Overall, when B atoms are introduced substitutionally into the network, the Fermi level shifts toward the valence edge. In configuration (1), (2), (4) and (7), all B atoms are four-fold and prefer to form three shorter bonds and one longer bond with Si. As concentration increases, more valence tail states are formed, and states move into the gap. Where B dimers or clusters are concerned, it seems that B-B bonds won't impact the doping as long as B atoms are four-fold as shown in EDOS of configuration (3). However, when more B dimers or clusters are formed, additional defect states appear near the conduction tail and they clutter the gap, as shown in configuration (5) and (6). Those defect states are associated with under- and over-coordinated Si atoms.

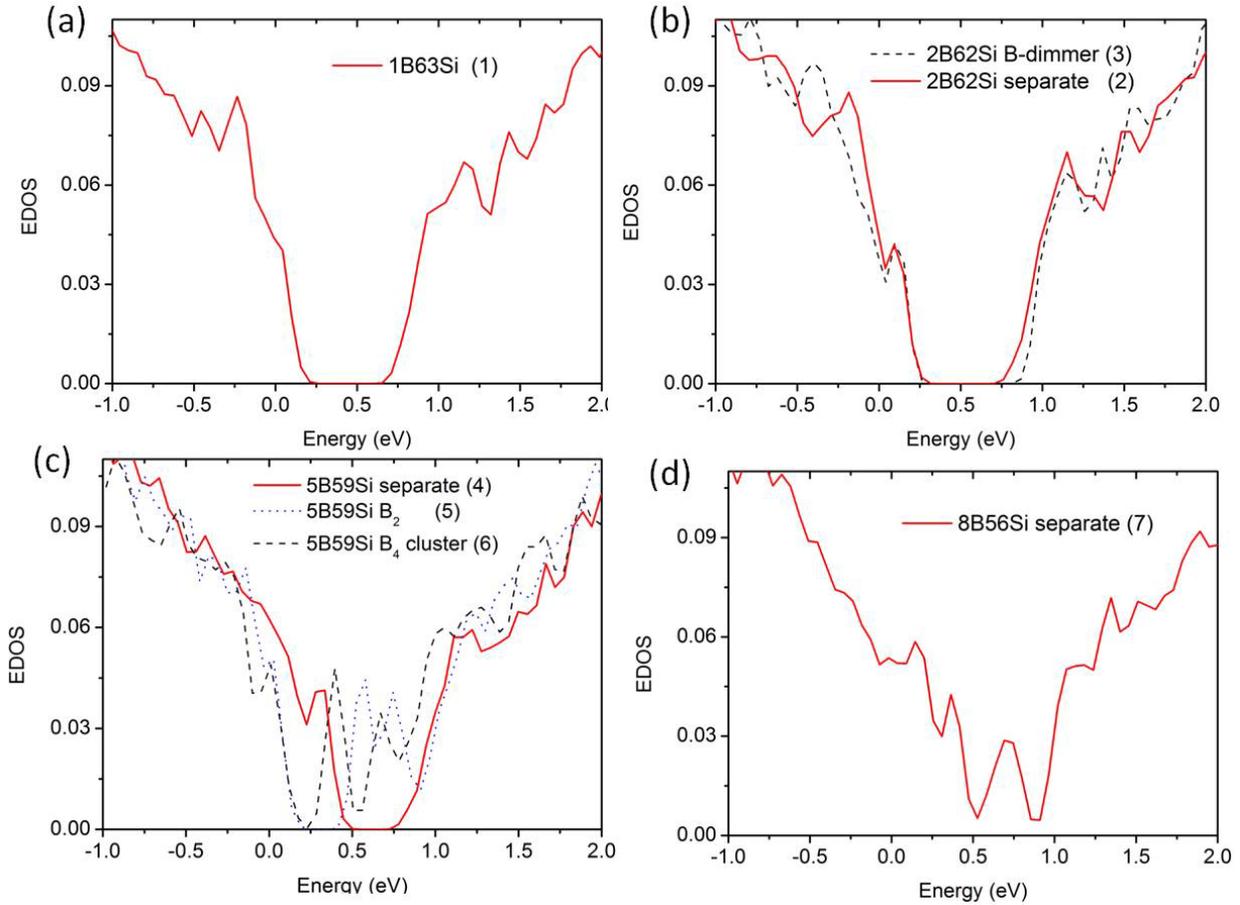

**Fig. 2** Boron doped a-Si with different concentrations. (a) 1.6% (b) 3.1% (c) 7.8% (d) 12.5%. In configuration (1), (2), (4) and (7), all B are bonded with four Si atoms. One B dimer is formed in configuration (3). Two B dimmers exist in configuration (5) and one $B_4$ cluster formed in configuration (6). The Fermi level is at 0 eV.

We conclude that tetrahedral B makes the Fermi level shift from mid-gap into the valence band tail and effectively dopes the system. However, as B concentration increase, more valence tail states are formed. B clusters may introduce tail and mid-gap states, which impact the electronic gap.

**Phosphorus doped a-Si**

We plot the EDOS of eight models [(1)-(8)] of P-doped a-Si in Fig. 3. We found that tetrahedral P forms deep donor states and the Fermi level shifts toward the conduction band tail. As P concentration increases, more defect states appear, and the gap closes (configurations (1), (2), (5), and (8)). P dimers and clusters dope the system so long as all P are four-fold (Configuration (3),(5)), but they also lead to defects which give rise to tail states. If P atoms are three-fold (configuration (4)), the configuration is non-doping.

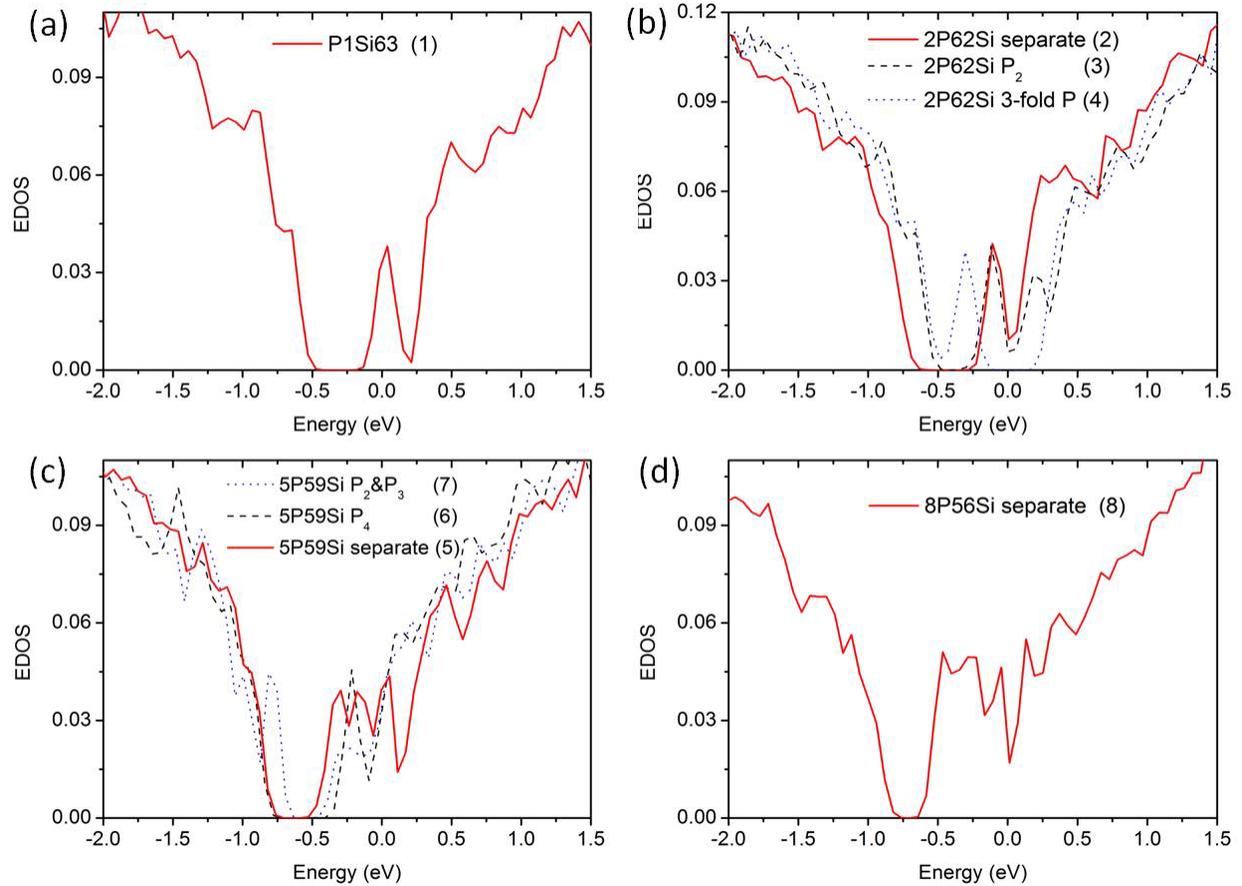

**Fig. 3** Phosphorus doped a-Si with different P concentration. (a) 1.6% (b) 3.1% (c) 7.8% (d) 12.5%. In (b) configuration(4) is non-doping configuration. The Fermi level is at 0eV.

Thus, for P doping, tetrahedral P dopes the system by shifting the Fermi level into the conduction tail. As for B doping, as concentration increases, more defect states move into gap. Three-fold P is a non-doping configuration.

**H passivation in B doped a-Si:H**

We report first that B (3,1) (B atoms bonded with three Si atoms and one H atom) is an effective doping configuration. The structural evolution and corresponding EDOS is plotted in Fig. 4 and Fig. 5.

The H is initially attached to a B atom and makes B form a metastable unit B(4,1) as in Fig. 4 (a). After relaxation, H breaks one Si-B bond forming a B(3,1) structure and leaves one Si DB, as in Fig. 4 (b). The EDOS of this configuration reveals that the Fermi level shifts back into gap. One mid-gap state forms due to the Si DB. However, if another H passivates the Si DB, like Fig.4 (c), the Fermi level shifts into valence band tail once again and the mid-gap state disappears. Thus, we conclude that the B(3,1) without a Si DB is an effective doping conformation. This confirms the simulation results reported in Ref [4].

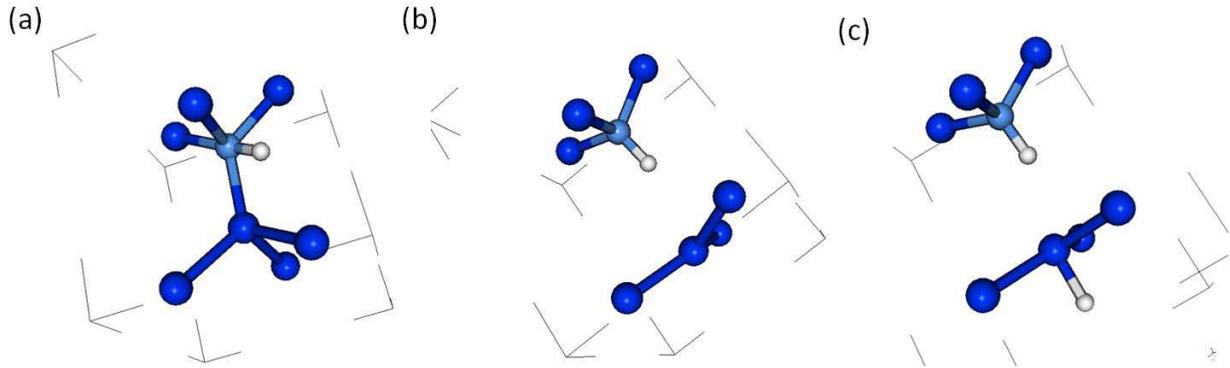

**Fig. 4** H passivation at B site. (a) H initially bond with B and B forms B(4,1) structure. (b) After relaxation, B becomes B(3,1) and leaves on Si DB. (c) Another H passivate the Si DB. The cyan (light,big) atom is B, blue(dark,big) atoms are Si and white(light,small) atom is H.

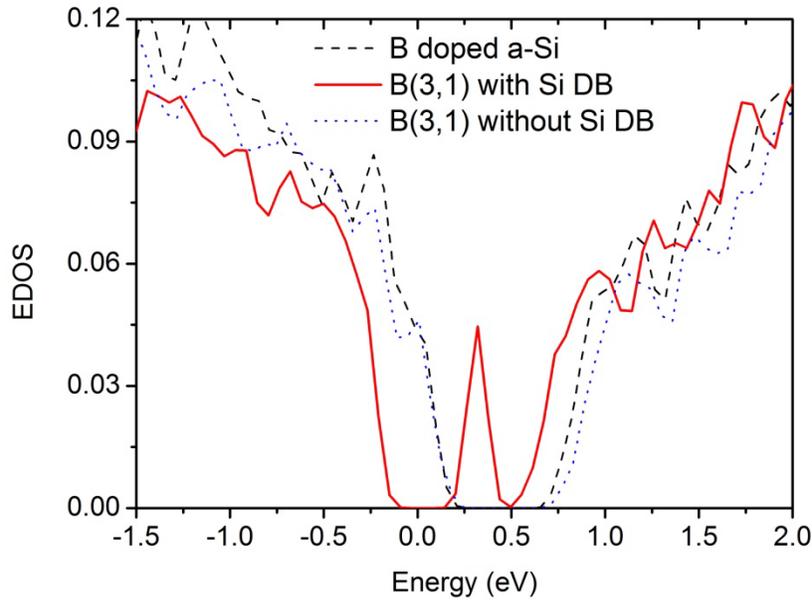

**Fig. 5** EDOS comparison of configurations during H passivation. The black (dashed) line shows the original B doped a-Si. The red(solid) line shows the EDOS of B(3,1) and one Si DB. The blue (dotted) line shows the EDOS when Si DB is passivated by another H. The Fermi level is at 0eV.

Next, we show that H

prefers to stay at the bond center (BC) when near a B atom and this BC H kills the doping. We show two cases of H passivation with different H-B distances in Fig.6. The top panel (a) of Fig. 6 shows the situation in which H is initially bonded to a Si neighbor of the B atom. After relaxation, H breaks the Si-Si bond and stays at the BC forming a B-Si-H-Si structure, the top panel [6(b)]. The EDOS shows that the Fermi level shifts back from the valence tail into the gap, indicating that the BC H kills the doping. The bottom panel of Fig.6 shows similar passivation with H bonded to a second neighbor Si of B atom and finally forming a B-Si-Si-H-Si structure. The EDOS of the B-Si-Si-H-Si structure shows that it is also a non-doping configuration. These results indicate that BC H sufficiently near a B atom will neutralize the doping. Notice that, in all cases, there is no reconstruction of B atoms -- which are still 4-fold after relaxation. There is no Si DB left in the network and no defect states in the gap. Further calculations indicate that there is an "H kill range" for BC H passivation. If the distance between H and B is beyond about 6.0 Å, the passivation may not occur.

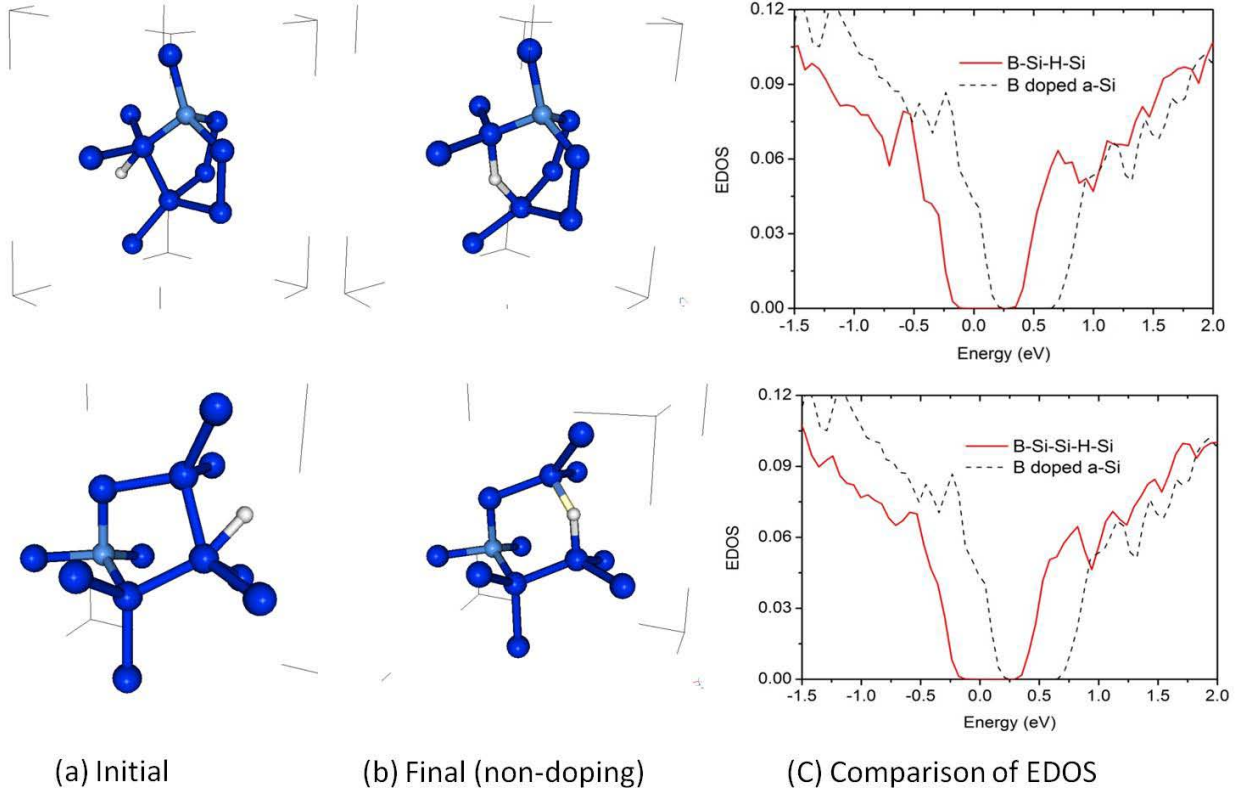

(a) Initial    (b) Final (non-doping)    (C) Comparison of EDOS

**Fig. 6** BC H passivates the doping. Top panel is the situation for B-Si-H-Si; bottom panel is the case for B-Si-Si-H-Si. The EDOS show the comparison between B doped a-Si and the final relaxation result when BC H exists in the network. The Fermi level is at 0eV.

The mechanism of this BC H passivation at low B concentration may be related to charge interactions. In Ref [2], at low concentration of B, it is confirmed that holes could be trapped at strained Si-Si bond centers. Thus, the H may compensate by staying at the BC and killing the doping structure. Another calculation in a-Si:H also confirm this. In a-Si:H, the Si (4,1) structure (Si bonded with four Si and one H) is stable. However, if one electron is removed from the system, H tends to break a Si-Si bond and occupy the BC position. Considering the "H poisoning range", it may be related to the exciton radius [6], which is about 5.9 Å in a-Si.

**H passivation in P doped a-Si:H**

In analogy with H passivation in B doped a-Si:H, we investigate H passivation in P-doped a-Si:H. We first report that P(3,1) is an effective doping configuration. The simulation is shown in Fig.7 and Fig.8. H is originally bonded to P forming a P(4,1) metastable structure. After relaxation, a P-Si bond breaks. H sticks to the P, forming a P(3,1) structure with a vestigial Si DB. The corresponding EDOS indicates that the deep donor state disappears, the Fermi level shifts back to the gap, and there is one mid-gap state formed (due to the Si DB). The configuration becomes non-doping. However, if another H passivates the Si DB as shown in Fig.7 (c), the Fermi level again shifts to the conduction tail. Thus, we conclude that P(3,1) is an effective doping structure, but the Si DB in the network may kill the doping.

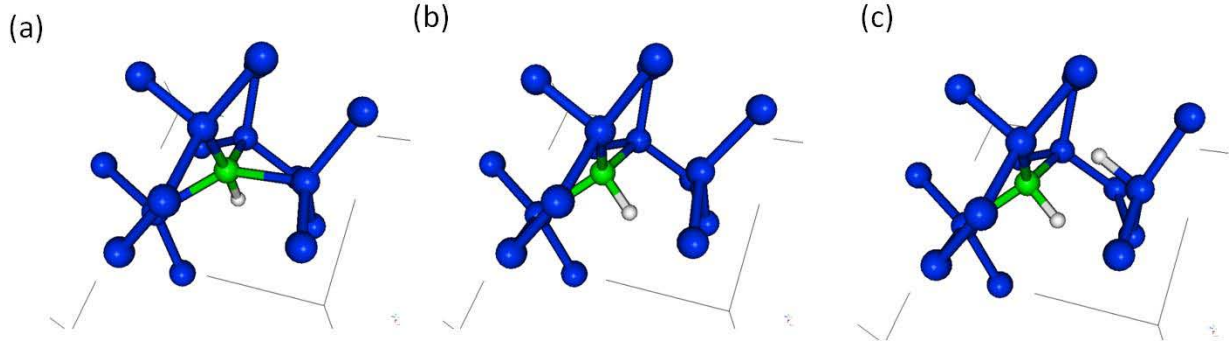

**Fig.7** H passivation at P site. (a) H bond with P forming a P(4,1) structure. (b) H breaks one P-Si bond and makes P form P(3,1) with one Si DB. (c) another H passivates the Si DB. Green (light,big) atom is P. Blue(dark,big) atom is Si. White(small) atoms is H.

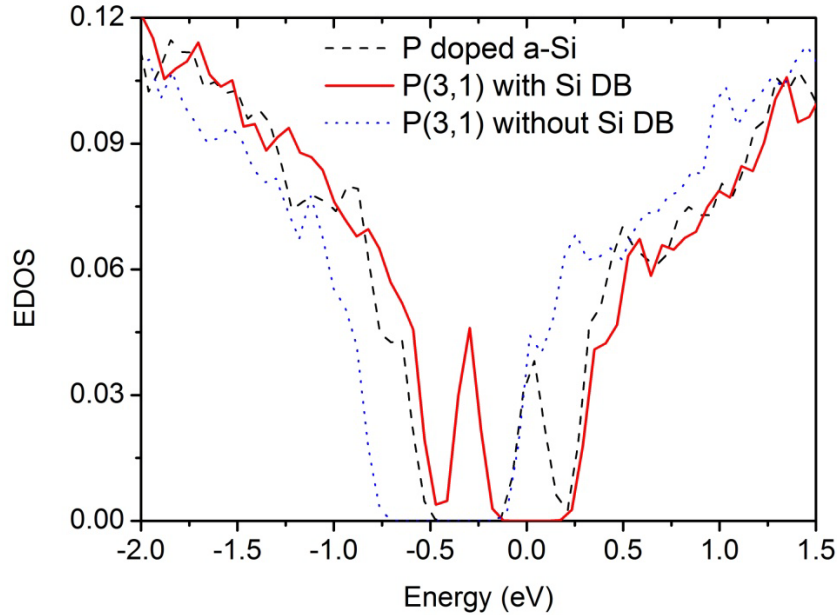

**Fig. 8** EDOS comparison of configurations during H passivation. Black (dashed) line is P doped a-Si. Red (solid) line is P(3,1) with one Si DB. Blue (dotted) line is the P(3,1) with another H passivate the Si DB. The Fermi level is at 0eV.

We then show two cases of H passivation in P doped a-Si:H in Fig.9. Unlike H passivation in B-doped a-Si:H, H in this network does not prefer BC position. In the top panel (a) of Fig.9, H is originally at BC of a P-Si bond and makes P(3,1) structure. However, after relaxation, the H-P bond ruptures, the H bonds with Si, and P becomes three-fold. The EDOS becomes non-doping as the deep donor state disappears and the Fermi level is now in the gap. The bottom panel of Fig. 9 shows another case. H is initially at a bond center of Si-Si and forms a P-Si-H-Si structure. After relaxation, the network reconstructs as P becomes three-fold and H sticks to a Si DB. The doping is rendered inactive. Notice that there are no defect states in the EDOS of final configuration. Further calculations show that again there exists an "H kill range"

(~ 6.0 Å). When H is sufficiently near a P site in a BC configuration, the network will reconstruct so that P becomes three-fold, H sticks to Si DB, and neutralizes the doping structure.

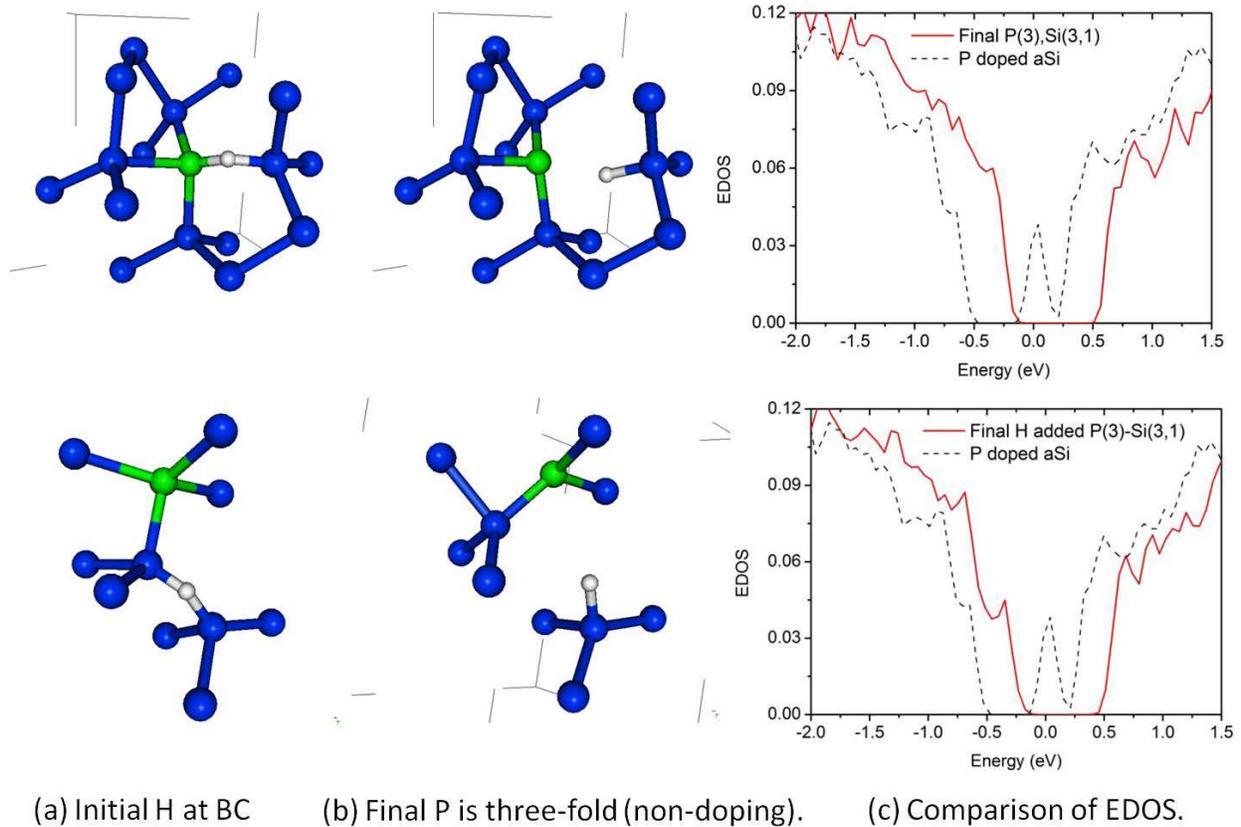

(a) Initial H at BC    (b) Final P is three-fold (non-doping).    (c) Comparison of EDOS.

**Fig. 9** H passivation in P doped a-Si:H. Top panel, H originally formed P-H-Si; bottom panel, H originally formed p-Si-H-Si. After relaxation, in both cases, P becomes three fold and non-doping configurations. Green (light, big) atom is P. Blue(dark, big) atom is Si. White (light, small) atom is H. The Fermi level is at 0eV.

Unlike H passivation for B, H does not prefer the BC position, instead it prefers to bond with Si. This result is consistent with NMR, which implies that P often has a H neighbor around 2.6 Å away (*not* the first neighbor distance).

**CONCLUSIONS**

Tetrahedral B and P dope the system, but high concentration of impurities introduce mid-gap states. Clusters also create defect states in the gap. H passivation is a key to understanding doping efficiency. Covalently bonded H: B(3Si,1H), P(3Si,1H), Si(3Si,H) are effective doping states. There exists an "H kill range": for B doping: when H sufficiently near B, H breaks the bond and stays at the bond center; for P doping, H atoms bonds to Si and makes P three-fold.

Most calculations in this paper are based on direct relaxations. We are currently performing long MD runs to study the thermal stability of B, P tetrahedral structure in a-Si, the interaction between H and impurities at high temperatures in a-Si:H, and the H passivation in a-Si:H with high concentrations of impurities distributed in various ways. Those results will appear in the future papers.

**ACKNOWLEDGMENTS**

We thank Dr. W. Windl and Dr. S. Estreicher for helpful suggestions. We thank the ARO under MURI W91NF-06-2-0026 and NSF under DMR 09-03225 for supporting this work.